\documentclass[%
 aip,
 amsmath,amssymb,
 reprint,%
]{revtex4-1}

\usepackage{graphicx}
\usepackage{dcolumn}
\usepackage{bm}

\usepackage[utf8]{inputenc}
\usepackage[T1]{fontenc}
\usepackage{mathptmx}
\usepackage{etoolbox}
\usepackage[hidelinks]{hyperref}
\usepackage[utf8]{inputenc}
\makeatletter
\def\@email#1#2{%
 \endgroup
 \patchcmd{\titleblock@produce}
  {\frontmatter@RRAPformat}
  {\frontmatter@RRAPformat{\produce@RRAP{*#1\href{mailto:#2}{#2}}}\frontmatter@RRAPformat}
  {}{}
}%
\makeatother
\begin{document}

\preprint{AIP/123-QED}

\title[]{Experimental Observation of Temporally-Evolving Stochastic Vibration Patterns in a Vibrating Motor}

\author{Adhinarayan Naembin Ashok}
\thanks{These authors contributed equally to this work.}
\affiliation{Department of Electrical and Electronics Engineering, Birla Institute of Technology and Science, Pilani -- Dubai Campus, Dubai International Academic City, Dubai, UAE 345055} 

\author{Levita Kris}
\thanks{These authors contributed equally to this work.}
\affiliation{Department of Electrical and Electronics Engineering, Birla Institute of Technology and Science, Pilani -- Dubai Campus, Dubai International Academic City, Dubai, UAE 345055} 
\affiliation{Department of Mechanical Engineering, Birla Institute of Technology and Science, Pilani -- Dubai Campus, Dubai International Academic City, Dubai, UAE 345055} 
\author{Adarsh Ganesan}%
\email{adarsh@dubai.bits-pilani.ac.in}
\affiliation{Department of Electrical and Electronics Engineering, Birla Institute of Technology and Science, Pilani -- Dubai Campus, Dubai International Academic City, Dubai, UAE 345055}
\affiliation{Department of Mechanical Engineering, Birla Institute of Technology and Science, Pilani -- Pilani Campus, Vidya Vihar, Pilani, India 333033}

\date{\today}

\begin{abstract}

\end{abstract}

\maketitle

\begin{quotation}
This paper presents the observations of temporally evolving stochastic vibration patterns of a coin vibrating motor. Various voltages are applied to the coin vibrating motor, and the resulting vibrations are recorded using an accelerometer. Although an overall upward trend in mean vibration amplitude is observed with increasing drive voltage, instantaneous waveforms displayed pronounced nonlinear and quasiperiodic amplitude modulations, frequency shifts, and stochastic deviations that intensified at higher voltages. Additional experiments involving periodic pressing of the motor and the propagating medium revealed the dependence of nonlinear electromechanical responses on the initial conditions. These results demonstrate that the dynamic behaviour of the coin-type motor is governed by a complex nonlinear dependence on current and displacement, with significant implications for precision control in miniature actuator applications.
\end{quotation}

Vibrating motors are actuators that convert electrical energy into mechanical motion\cite{app10248915}. These motors are electromagnetically coupled with an unbalanced mass \cite{Osadchyy2023}$^,$\cite{DONG2016134}. Here, electromagnetic forces generated by a coil and magnet induce rotational motion, which in turn is converted into vibration through an intentionally unbalanced mass\cite{XU1994663}$^,$\cite{kim2019analysis}. Advances in actuator design and materials are pushing vibrating motor technologies toward higher bandwidths. This enables the device to have more precise control \cite{Wang2022}$^,$\cite{9795333}$^,$\cite{refId0}$^,$\cite{mertens2011individual}. 

Vibrating motors have found their application in miniature fans and actuators to dissipate heat from densely packed electronics, thereby stabilizing imaging quality and ensuring operational safety during prolonged usage \cite{puangmali2011state}$^,$\cite{choi2013vibrotactile}. More recently, vibrating motors have found use in wearable devices where they mechanically stimulate target muscle groups \cite{popovic2005nms}. These can also be integrated into a portable diagnosis and prognosis system for selected diseases \cite{7470536}$^,$\cite{9442056}$^,$\cite{4415100}. Beyond core medical infrastructure, vibrating motors have become indispensable in rehabilitation medicine and physiotherapy, particularly for elderly populations and for healing sports-related injuries\cite{10010851}. Portable vibration-based physiotherapy tools, like percussive massagers and vibration foam rollers, now widely incorporate linear resonant actuators (LRAs) or eccentric rotating mass (ERM) motors due to their precision control, low power demand, and durability \cite{puangmali2011state}$^,$\cite{lien2020foam}. In the field of metallurgy, vibrating motors are very effective in bulk material processing and characterising \cite{Gursky2022}$^,$\cite{mi11080753}. 

In consumer devices such as cell phones, handsets and pagers, compact coreless DC motors are used \cite{1519521}$^,$\cite{Chen2013}. These motors have evolved from being an element for simple alert mechanism to being a sophisticated interface element, playing a vital role in enhancing functionality, accessibility, and user engagement across a broad spectrum of modern electronics. Their ability to provide silent and efficient feedback enhances user interaction, particularly in smartphones, smartwatches, fitness trackers, and gaming controllers. As haptic technology continues to advance, with improvements in motor types such as linear resonant actuators and eccentric rotating mass motors, vibration feedback is becoming increasingly precise and dynamic \cite{DONG2016134}$^,$\cite{Wang2022}$^,$\cite{lien2020foam}. This not only boosts functionality but also expands the potential applications of vibrating motors in future consumer devices. Ultimately, their small size, low power requirements, and versatility make them indispensable in the design of modern user-centric technologies \cite{Chen2013}.

Although the non-linear dynamics of the vibrating motor has been studied before in the context of individual and collective behavior \cite{mertens2011individual}, this paper showcases an interesting stochastic response of a coin vibrating motor. When this vibrating motor is driven at a particular voltage, a temporally evolving stochastic vibration pattern is observed. While the oscillatory patterns associated with the vibration change with time, their amplitude stayed almost constant \cite{mi11080753}. Upon increasing the voltage, the amplitude of vibrations monotonously increased. While this trend of monotonous increase of amplitude with voltage is consistent across different runs, the actual values of amplitude did vary. This paper presents these observations and explores the related mechanisms.

\begin{figure*}
\centering
\includegraphics[width=11.8cm,height=11cm]{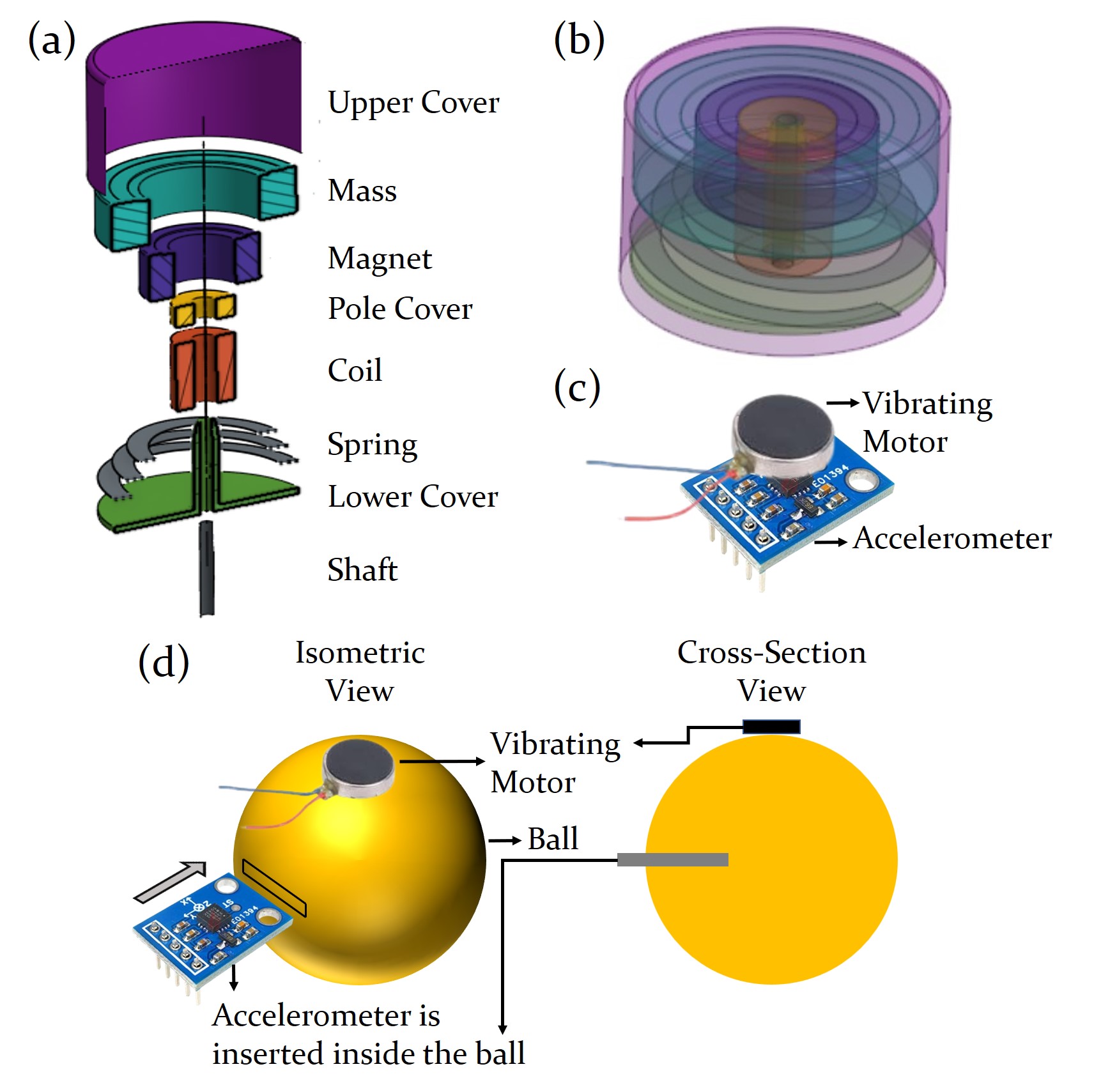}
\caption{\label{Device} (a) An exploded isometric view showing the internal assembly of a coin-type vibrating motor, (b) A semi-transparent 3-D visualization of the assembled motor with internal component visibility, (c) Setup for recording the vibrations of vibrating motor using accelerometer, (d) Setup for accelerometer measurements of vibrations in a ball launched using a vibrating motor.}
\end{figure*}

\textit{Device Architecture}--The coin vibrating motor consists of a weight, a ring motor, a rotor with commutation points attached in the front, coils assembled at the back, and a ring magnet \cite{kim2019analysis}$^,$\cite{Chen2013}(Figs. 1(a) and 1(b)). A vibrating motor functions based on an unbalanced centrifugal force that causes the motor to wobble and vibrate \cite{DONG2016134}. The vibration amplitude and direction depend on the mass of the weight, its distance from the shaft, and the motor’s rotational speed \cite{app10248915}$^,$\cite{Chen2013}. Increasing the motor's voltage raises its speed, thereby amplifying both the vibration frequency and amplitude \cite{DONG2016134}$^,$\cite{9795333}. 

\textit{Experimental Method}--The  coin vibrating motor is operated at different voltages set using the Arduino UNO code. The vibrations are recorded using an ADXL 335 accelerometer (Fig. 1(c)). While the accelerometer records vibrations in x-,y- and z-axes, the vibration amplitudes along x-axis are significantly higher than that of y-axis and z-axis. The readings are taken in the form of analog-to-digital converter (ADC) signals, and interesting vibration patterns are henceforth observed. For each operating condition, we conducted three runs to assess the reproducibility of vibration patterns. In order to analyze the effect of the propagation medium on the vibration characteristics of the vibrator, we attached the vibrating motor to a foam stress ball as shown in Fig. 1(d). A slit is created within this ball, and the accelerometer is inserted inside to record vibrations.

\begin{figure*}
\centering
\includegraphics[width=17cm,height=8.5cm]{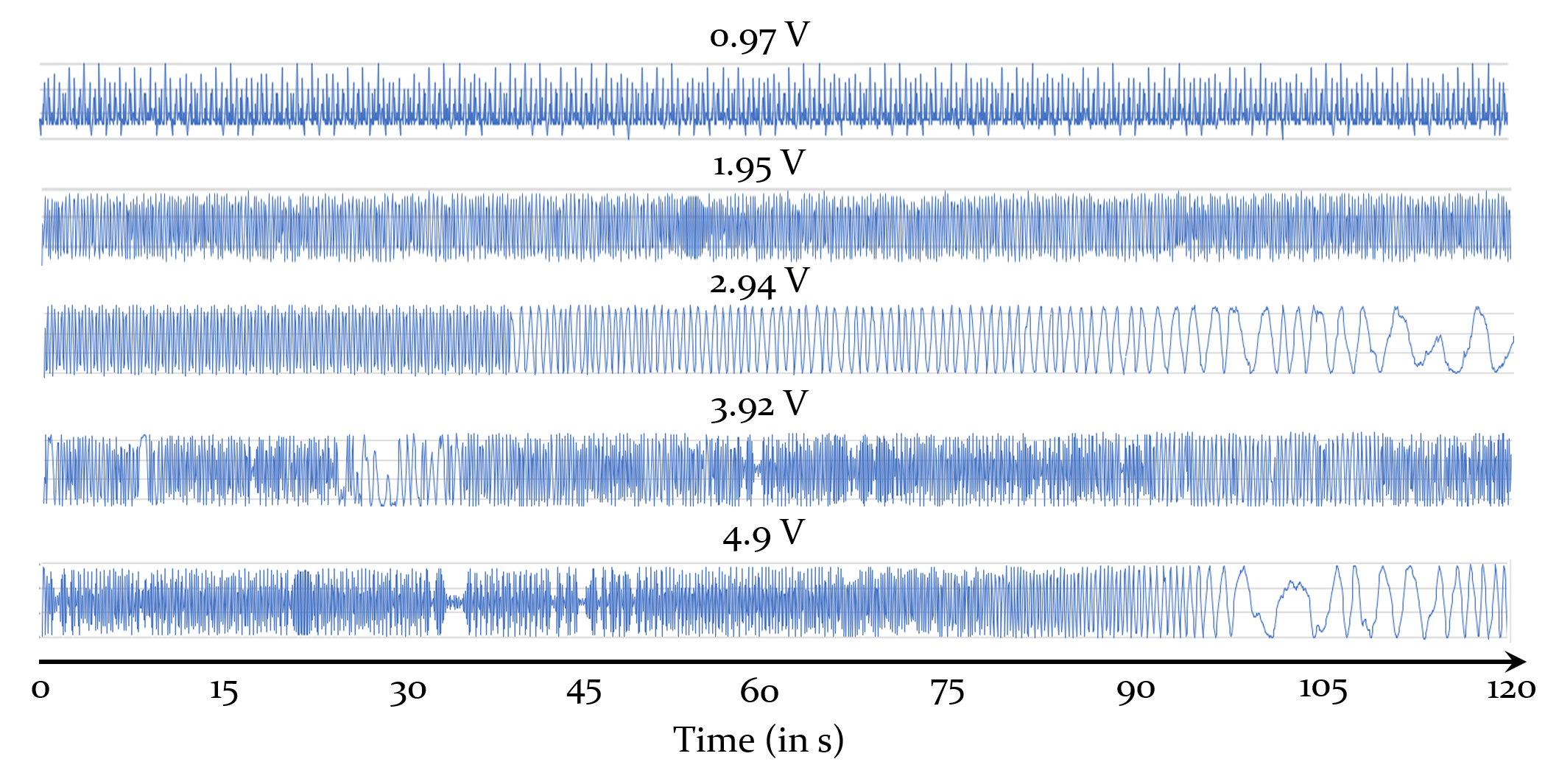}
\caption{\label{Observe1} Stochastic vibration responses along the x-axis measured for different input voltages}
\end{figure*}

\begin{figure}
\centering
\includegraphics[width=7cm,height=5cm]{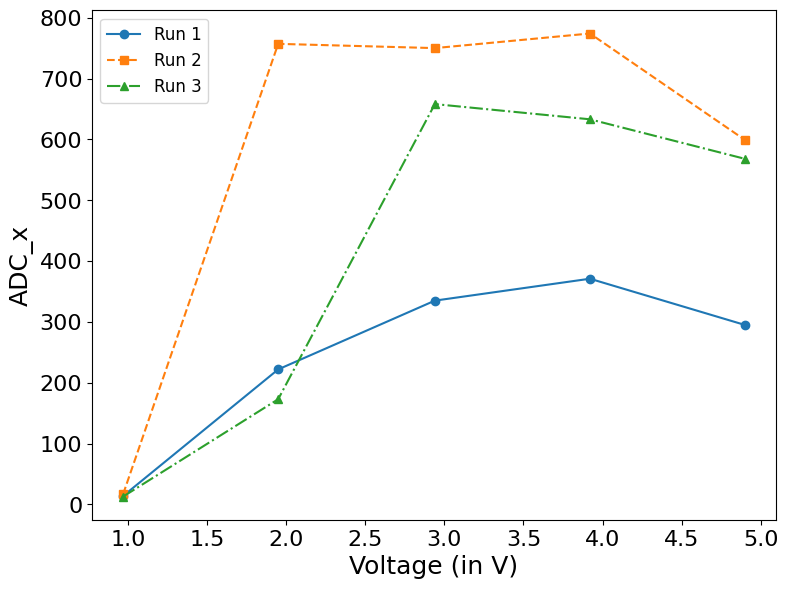}
\caption{\label{Observe2} The vibration amplitude plotted against input voltage across multiple experimental runs, highlighting variations observed under consistent test conditions}
\end{figure}

\textit{Coupled Dynamics}-- Using Maxwell's equations, the magnetic flux ($\phi$), inductance (L) and speedance (K) are obtained as a function of displacement $x$ and current $i$ \cite{kim2019analysis}.
\begin{equation}
\begin{array}{l}
\phi(x, i) = NA\vec{B}(x, i) \\[8pt]
L(x, i) = \dfrac{\Delta \phi(x, i)}{\Delta i} \\[8pt]
K(x, i) = \dfrac{\Delta \phi(x, i)}{\Delta x}
\end{array}
\end{equation}
Here, $N$, $A$, and $\vec{B}$ are coil turns, coil sectional areas, and average flux density, respectively. The interactions between electromagnetic forces and mechanical momentum can be formulated using Maxwell stress tensor as denoted in (2). 
\begin{equation}
F_m = \frac{1}{\mu_0} \iint_A \left[ \left( \vec{B} \cdot \vec{n} \right) \vec{B} - \frac{1}{2} \vert \vec{B} \vert ^2 \vec{n}\right] dA
\end{equation}
Here, $F_m$ is the magnetic force, $\vec{n}$ is the unit normal vector, and $A$ is the surface under consideration. The total force $F_{total}$ arises due to the presence of a permanent magnet, magnetic materials, and electrical current, and is comprised of two components: cogging force ($F_{\text{cogging}}$) and current-induced force ($F_{\text{current}}$). Cogging force is an attractive force between the material and permanent magnet. To determine the force generated by the current, cogging force is subtracted from the overall force acting on the vibrating element. The force factor $f(y,i)$ is then calculated by taking the ratio of this current-induced force to the current.
\begin{equation}
\begin{aligned}
F_{\text{cogging}}(x) &= F_{\text{total}}(x, i)\big|_{i=0} \\
F_{\text{current}}(x, i) &= F_{\text{total}}(x, i) - F_{\text{cogging}}(x) \\
f(y, i) &= \frac{F_{\text{current}}(x, i)}{i}
\end{aligned}
\end{equation}

The force factor, along with the other parameters inductance and speedance, exhibits nonlinear behavior, since they depend on time-varying current and displacement. This is due to the complex electromagnetic interactions within the system of the coin vibrating motor. The nonlinearity appears in the voltage expression as:
\begin{equation}
V = iR + L(x, i) \frac{di}{dt} + K(x, i) \frac{dx}{dt}
\end{equation}
where, $V$ and $R$ are the applied voltage and the resistance of the device, respectively. The displacement amplitudes of vibrations $x(t)$ can thus be obtained from (4) for any applied voltage $V$. The nonlinear interactions present in (4) can result in stochastic behaviour of $x(t)$ with extreme sensitivity to initial conditions.

\begin{figure*}
\centering
\includegraphics[width=17cm,height=9cm]{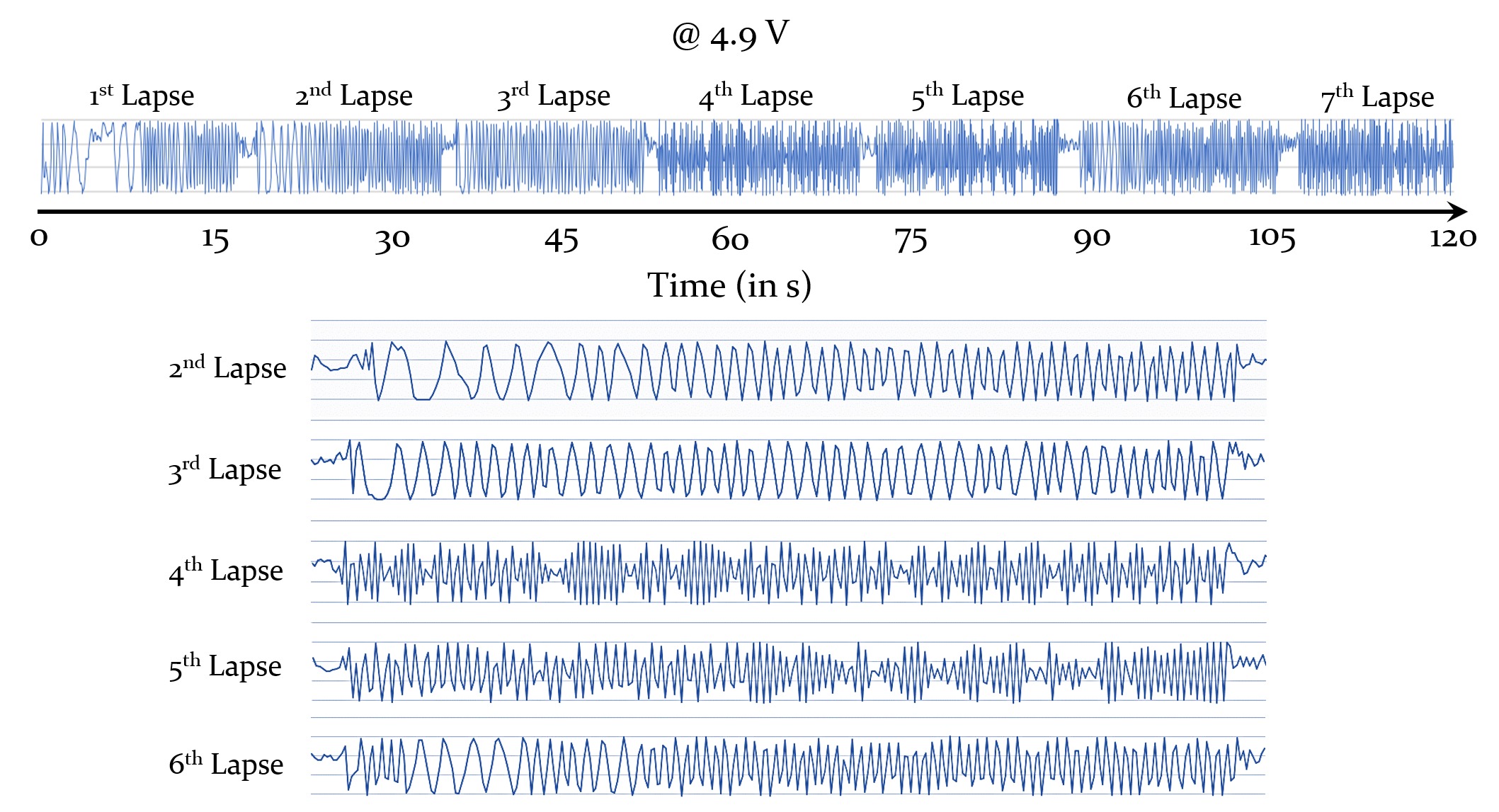}
\caption{\label{Observe3} Stochastic vibration response recorded as the vibrator is periodically pressed at intervals of 20 seconds}
\end{figure*}

\begin{figure*}
\centering
\includegraphics[width=17cm,height=9cm]{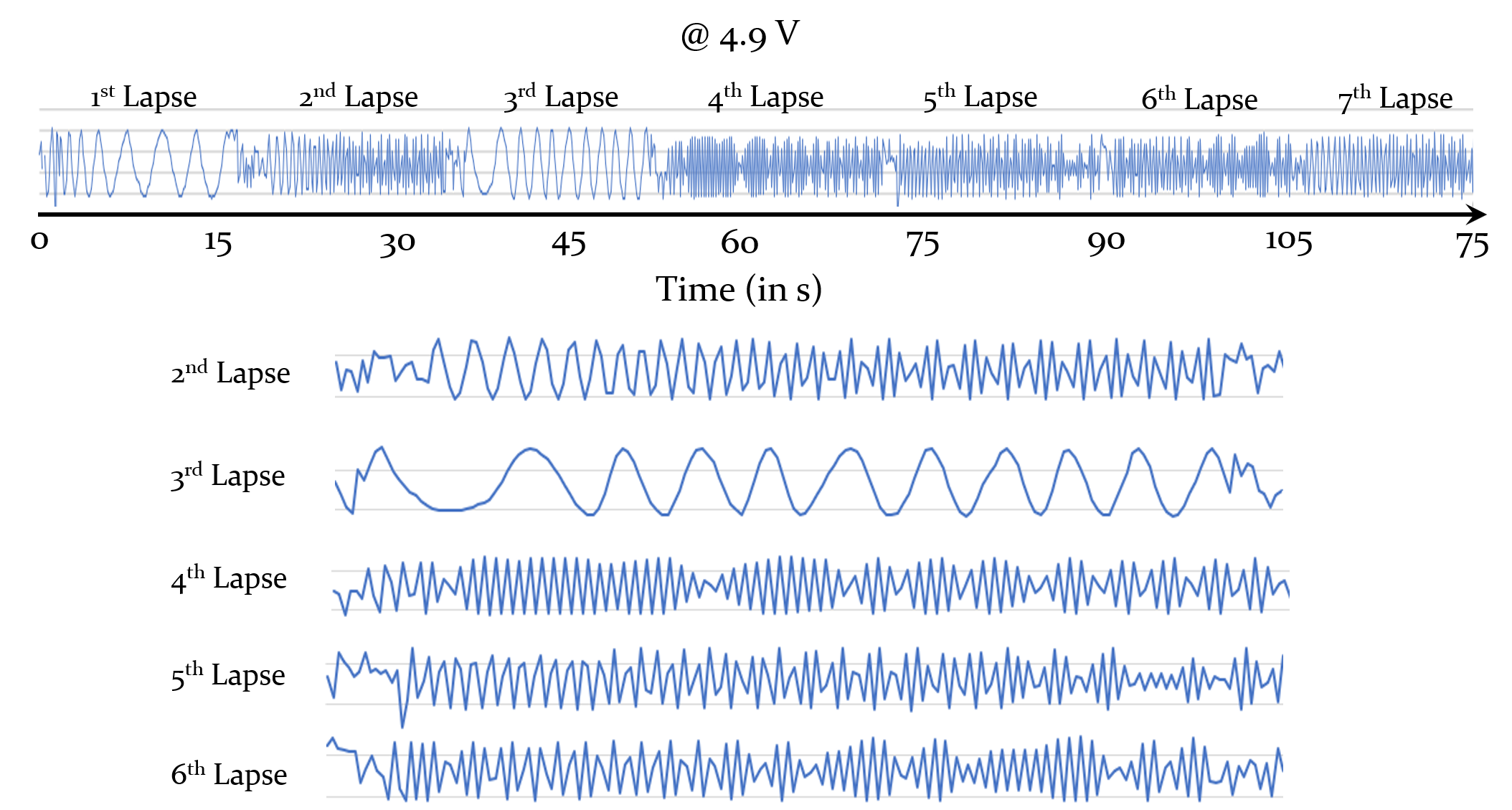}
\caption{\label{Observe4} Stochastic vibration response recorded as the ball is periodically pressed at intervals of 20 seconds}
\end{figure*}

\textit{Observations}-- Fig. 2 presents the stochastic vibration responses of the coin-type motor. At low enough voltage of 0.97 V, the vibrating motor does not vibrate significantly. The accelerometer records the background vibration patterns at this voltage. These vibrations suffer from negligible stochastic deviations. However, as the applied voltage is increased above a threshold, the vibrating motor starts to vibrate at significant amplitudes. At 1.95 V, we observe quasiperiodic amplitude modulations coupled with both abrupt and gradual shifts in oscillation frequency. Under 2.94 V excitation, the mean oscillation frequency gradually decays over the test duration with pronounced amplitude irregularities arise after 105 s. At 3.92V, reduced amplitude excursions occur near 30s and briefly at 60s, although these anomalies are not sustained throughout the entire time domain of the experiment. There also exists persistent beating patterns throughout the time domain of the experimental run. Finally, at 4.9V, the vibration trace exhibits a progressive rarefaction with intermittent beating patterns across the time domain. Collectively, these observations demonstrate that the inherent stochastic dynamics of the coin-vibrating motor resulting from the underlying nonlinear electromechanical coupling \cite{refId0}.

Fig. 3 presents the run-to-run variation observed in the readings of vibration amplitudes. In all three trials, the temporal waveform closely resembles the random/stochastic vibration patterns shown in Fig. 2. Also, in all runs, the vibration amplitudes increase with voltage before saturating at higher voltages. Although the trends in these runs are similar, the actual amplitude values vary across runs.

Now, in order to prove the dependence of the observed stochastic vibration patterns on the initial conditions, the vibrating motor is pressed at regular time intervals. By pressing the vibrating motor, the vibrations are stopped. Upon releasing it, the vibrations are restarted. After each press, the initial conditions for the vibrating motor can thus change. Henceforth, the resulting dynamics of (4) also changes, causing the vibration patterns to differ. Fig. 4 presents the results of this experiment. Here, 4.94 V is applied to the vibrating motor. The motor is pressed at a rough interval of 20 seconds. After each press, the time-domain waveform of the vibrations begins to change, rather than being consistent, which is evident from the waveforms corresponding to different lapses. This hence proves our assertion.

In order to now study the effect of propagation medium on the observed stochastic behavior, the vibrating motor is attached to a foam stress ball. The accelerometer is inserted inside a small slit cut out of the ball. Fig. 5. shows the vibration responses corresponding to this experiment. Here again, 4.9V is the applied voltage. The ball is pressed periodically after every 20 seconds. Very much like the previous experiment, the vibration waveform evolves continuously over the duration of the test. However, by comparing Figs. 4 and 5, it can be inferred that the vibrations demonstrate lower vibration amplitudes and frequencies in the presence of a medium. Nonetheless, the inherent stochastic nature of vibrating motor is preserved. This thus enables the possibility of using stochastic motion of vibrating motors for probing mechanical properties of target medium.


\textit{Conclusions}-- The observations demonstrate that the vibration behavior of the coin-type motor arises from complex nonlinear electromechanical interactions \cite{XU1994663}. Instantaneous acceleration profiles reveal unpredictable amplitude and frequency behaviors in the observed waveform even under steady voltage drives in different runs. This indicates internal dynamic instabilities and electromechanical coupling. The application of mechanical loads induces residual deformation in the surrounding medium, further disturbing the response of the motor by possibly exciting other vibration modes and reshaping the waveform of the observed pattern\cite{ganesan2016observation}$^,$\cite{ganesan2017excitation}$^,$\cite{ganesan2017phononic}. Therefore, the coin motor behaves as a nonlinear electromechanical oscillator, exhibiting amplitude and frequency modulations alongside stochastic fluctuations. Deformations introduce parametric disturbances that cause randomness in its vibration patterns making them temporally-evolving stochastic vibrations. Future work will involve fundamental investigations of nonlinear dynamics of vibrating motors\cite{mertens2011individual}$^,$\cite{ganesan2016observation}$^,$\cite{ganesan2017excitation}$^,$\cite{ganesan2017phononic}. Besides this, the possibility of probing mechanics of biological tissues will also be explored.

\begin{acknowledgments}
The authors sincerely thank the School of Interdisciplinary Research and Entrepreneurship (SIRE), BITS Pilani, for providing the funding support for this project through its SPARKLE program. 
\end{acknowledgments}

\section*{Data Availability Statement}
The data that support the findings of this study are available from the corresponding author upon reasonable request.

\nocite{*}
\bibliography{ref}

\begin{thebibliography}{24}%
\makeatletter
\providecommand \@ifxundefined [1]{%
 \@ifx{#1\undefined}
}%
\providecommand \@ifnum [1]{%
 \ifnum #1\expandafter \@firstoftwo
 \else \expandafter \@secondoftwo
 \fi
}%
\providecommand \@ifx [1]{%
 \ifx #1\expandafter \@firstoftwo
 \else \expandafter \@secondoftwo
 \fi
}%
\providecommand \natexlab [1]{#1}%
\providecommand \enquote  [1]{``#1''}%
\providecommand \bibnamefont  [1]{#1}%
\providecommand \bibfnamefont [1]{#1}%
\providecommand \citenamefont [1]{#1}%
\providecommand \href@noop [0]{\@secondoftwo}%
\providecommand \href [0]{\begingroup \@sanitize@url \@href}%
\providecommand \@href[1]{\@@startlink{#1}\@@href}%
\providecommand \@@href[1]{\endgroup#1\@@endlink}%
\providecommand \@sanitize@url [0]{\catcode `\\12\catcode `\$12\catcode `\&12\catcode `\#12\catcode `\^12\catcode `\_12\catcode `\%12\relax}%
\providecommand \@@startlink[1]{}%
\providecommand \@@endlink[0]{}%
\providecommand \url  [0]{\begingroup\@sanitize@url \@url }%
\providecommand \@url [1]{\endgroup\@href {#1}{\urlprefix }}%
\providecommand \urlprefix  [0]{URL }%
\providecommand \Eprint [0]{\href }%
\providecommand \doibase [0]{http://dx.doi.org/}%
\providecommand \selectlanguage [0]{\@gobble}%
\providecommand \bibinfo  [0]{\@secondoftwo}%
\providecommand \bibfield  [0]{\@secondoftwo}%
\providecommand \translation [1]{[#1]}%
\providecommand \BibitemOpen [0]{}%
\providecommand \bibitemStop [0]{}%
\providecommand \bibitemNoStop [0]{.\EOS\space}%
\providecommand \EOS [0]{\spacefactor3000\relax}%
\providecommand \BibitemShut  [1]{\csname bibitem#1\endcsname}%
\let\auto@bib@innerbib\@empty
\bibitem [{\citenamefont {Jiang}, \citenamefont {Park},\ and\ \citenamefont {Hwang}(2020)}]{app10248915}%
  \BibitemOpen
  \bibfield  {author} {\bibinfo {author} {\bibfnamefont {Z.-X.}\ \bibnamefont {Jiang}}, \bibinfo {author} {\bibfnamefont {K.-H.}\ \bibnamefont {Park}}, \ and\ \bibinfo {author} {\bibfnamefont {S.-M.}\ \bibnamefont {Hwang}},\ }\href {\doibase 10.3390/app10248915} {\bibfield  {journal} {\bibinfo  {journal} {Applied Sciences}\ }\textbf {\bibinfo {volume} {10}} (\bibinfo {year} {2020}),\ 10.3390/app10248915}\BibitemShut {NoStop}%
\bibitem [{\citenamefont {Osadchyy}\ \emph {et~al.}(2023)\citenamefont {Osadchyy}, \citenamefont {Nazarova}, \citenamefont {Hutsol}, \citenamefont {Glowacki}, \citenamefont {Mudryk}, \citenamefont {Bryś}, \citenamefont {Rud}, \citenamefont {Tulej},\ and\ \citenamefont {Sojak}}]{Osadchyy2023}%
  \BibitemOpen
  \bibfield  {author} {\bibinfo {author} {\bibfnamefont {V.}~\bibnamefont {Osadchyy}}, \bibinfo {author} {\bibfnamefont {O.}~\bibnamefont {Nazarova}}, \bibinfo {author} {\bibfnamefont {T.}~\bibnamefont {Hutsol}}, \bibinfo {author} {\bibfnamefont {S.}~\bibnamefont {Glowacki}}, \bibinfo {author} {\bibfnamefont {K.}~\bibnamefont {Mudryk}}, \bibinfo {author} {\bibfnamefont {A.}~\bibnamefont {Bryś}}, \bibinfo {author} {\bibfnamefont {A.}~\bibnamefont {Rud}}, \bibinfo {author} {\bibfnamefont {W.}~\bibnamefont {Tulej}}, \ and\ \bibinfo {author} {\bibfnamefont {M.}~\bibnamefont {Sojak}},\ }\href {\doibase 10.3390/s23042170} {\bibfield  {journal} {\bibinfo  {journal} {Sensors}\ }\textbf {\bibinfo {volume} {23}},\ \bibinfo {pages} {2170} (\bibinfo {year} {2023})}\BibitemShut {NoStop}%
\bibitem [{\citenamefont {Dong}\ \emph {et~al.}(2016)\citenamefont {Dong}, \citenamefont {Yang}, \citenamefont {Chen}, \citenamefont {Xu}, \citenamefont {Meng},\ and\ \citenamefont {Ou}}]{DONG2016134}%
  \BibitemOpen
  \bibfield  {author} {\bibinfo {author} {\bibfnamefont {Z.}~\bibnamefont {Dong}}, \bibinfo {author} {\bibfnamefont {M.}~\bibnamefont {Yang}}, \bibinfo {author} {\bibfnamefont {Z.}~\bibnamefont {Chen}}, \bibinfo {author} {\bibfnamefont {L.}~\bibnamefont {Xu}}, \bibinfo {author} {\bibfnamefont {F.}~\bibnamefont {Meng}}, \ and\ \bibinfo {author} {\bibfnamefont {W.}~\bibnamefont {Ou}},\ }\href {\doibase 10.1016/j.ultras.2016.06.004} {\bibfield  {journal} {\bibinfo  {journal} {Ultrasonics}\ } (\bibinfo {year} {2016}),\ 10.1016/j.ultras.2016.06.004}\BibitemShut {NoStop}%
\bibitem [{\citenamefont {Xu}\ and\ \citenamefont {Marangoni}(1994)}]{XU1994663}%
  \BibitemOpen
  \bibfield  {author} {\bibinfo {author} {\bibfnamefont {M.}~\bibnamefont {Xu}}\ and\ \bibinfo {author} {\bibfnamefont {R.}~\bibnamefont {Marangoni}},\ }\href {\doibase https://doi.org/10.1006/jsvi.1994.1405} {\bibfield  {journal} {\bibinfo  {journal} {Journal of Sound and Vibration}\ }\textbf {\bibinfo {volume} {176}},\ \bibinfo {pages} {663} (\bibinfo {year} {1994})}\BibitemShut {NoStop}%
\bibitem [{\citenamefont {Kim}, \citenamefont {Jiang},\ and\ \citenamefont {Hwang}(2019)}]{kim2019analysis}%
  \BibitemOpen
  \bibfield  {author} {\bibinfo {author} {\bibfnamefont {J.-H.}\ \bibnamefont {Kim}}, \bibinfo {author} {\bibfnamefont {Y.-W.}\ \bibnamefont {Jiang}}, \ and\ \bibinfo {author} {\bibfnamefont {S.-M.}\ \bibnamefont {Hwang}},\ }\href@noop {} {\bibfield  {journal} {\bibinfo  {journal} {Applied Sciences}\ }\textbf {\bibinfo {volume} {9}},\ \bibinfo {pages} {1434} (\bibinfo {year} {2019})}\BibitemShut {NoStop}%
\bibitem [{\citenamefont {Wang}\ \emph {et~al.}(2022)\citenamefont {Wang}, \citenamefont {Dai}, \citenamefont {Zhang}, \citenamefont {Sayyad}, \citenamefont {Sugumar}, \citenamefont {Kumar},\ and\ \citenamefont {Asenso}}]{Wang2022}%
  \BibitemOpen
  \bibfield  {author} {\bibinfo {author} {\bibfnamefont {D.}~\bibnamefont {Wang}}, \bibinfo {author} {\bibfnamefont {L.}~\bibnamefont {Dai}}, \bibinfo {author} {\bibfnamefont {X.}~\bibnamefont {Zhang}}, \bibinfo {author} {\bibfnamefont {S.}~\bibnamefont {Sayyad}}, \bibinfo {author} {\bibfnamefont {R.}~\bibnamefont {Sugumar}}, \bibinfo {author} {\bibfnamefont {K.}~\bibnamefont {Kumar}}, \ and\ \bibinfo {author} {\bibfnamefont {E.}~\bibnamefont {Asenso}},\ }\href {\doibase 10.1049/tje2.12203} {\bibfield  {journal} {\bibinfo  {journal} {The Journal of Engineering}\ }\textbf {\bibinfo {volume} {2022}},\ \bibinfo {pages} {1124} (\bibinfo {year} {2022})},\ \Eprint {http://arxiv.org/abs/https://ietresearch.onlinelibrary.wiley.com/doi/pdf/10.1049/tje2.12203} {https://ietresearch.onlinelibrary.wiley.com/doi/pdf/10.1049/tje2.12203} \BibitemShut {NoStop}%
\bibitem [{\citenamefont {Xing}, \citenamefont {Wang},\ and\ \citenamefont {Zhao}(2022)}]{9795333}%
  \BibitemOpen
  \bibfield  {author} {\bibinfo {author} {\bibfnamefont {Z.}~\bibnamefont {Xing}}, \bibinfo {author} {\bibfnamefont {X.}~\bibnamefont {Wang}}, \ and\ \bibinfo {author} {\bibfnamefont {W.}~\bibnamefont {Zhao}},\ }\href {\doibase 10.1109/TTE.2022.3183074} {\bibfield  {journal} {\bibinfo  {journal} {IEEE Transactions on Transportation Electrification}\ }\textbf {\bibinfo {volume} {8}},\ \bibinfo {pages} {4337} (\bibinfo {year} {2022})}\BibitemShut {NoStop}%
\bibitem [{\citenamefont {{Toirov, Olimjon}}\ \emph {et~al.}(2023)\citenamefont {{Toirov, Olimjon}}, \citenamefont {{Khalikova, Malika}}, \citenamefont {{Jumaeva, Dilnoza}},\ and\ \citenamefont {{Kakharov, Sergey}}}]{refId0}%
  \BibitemOpen
  \bibfield  {author} {\bibinfo {author} {\bibnamefont {{Toirov, Olimjon}}}, \bibinfo {author} {\bibnamefont {{Khalikova, Malika}}}, \bibinfo {author} {\bibnamefont {{Jumaeva, Dilnoza}}}, \ and\ \bibinfo {author} {\bibnamefont {{Kakharov, Sergey}}},\ }\href {\doibase 10.1051/e3sconf/202340105089} {\bibfield  {journal} {\bibinfo  {journal} {E3S Web of Conf.}\ }\textbf {\bibinfo {volume} {401}},\ \bibinfo {pages} {05089} (\bibinfo {year} {2023})}\BibitemShut {NoStop}%
\bibitem [{\citenamefont {Mertens}\ and\ \citenamefont {Weaver}(2011)}]{mertens2011individual}%
  \BibitemOpen
  \bibfield  {author} {\bibinfo {author} {\bibfnamefont {D.}~\bibnamefont {Mertens}}\ and\ \bibinfo {author} {\bibfnamefont {R.}~\bibnamefont {Weaver}},\ }\href {\doibase 10.1002/cplx.20352} {\bibfield  {journal} {\bibinfo  {journal} {Complexity}\ }\textbf {\bibinfo {volume} {16}},\ \bibinfo {pages} {45} (\bibinfo {year} {2011})}\BibitemShut {NoStop}%
\bibitem [{\citenamefont {Puangmali}\ \emph {et~al.}(2011)\citenamefont {Puangmali}, \citenamefont {Althoefer}, \citenamefont {Seneviratne},\ and\ \citenamefont {Murphy}}]{puangmali2011state}%
  \BibitemOpen
  \bibfield  {author} {\bibinfo {author} {\bibfnamefont {P.}~\bibnamefont {Puangmali}}, \bibinfo {author} {\bibfnamefont {K.}~\bibnamefont {Althoefer}}, \bibinfo {author} {\bibfnamefont {L.~D.}\ \bibnamefont {Seneviratne}}, \ and\ \bibinfo {author} {\bibfnamefont {D.}~\bibnamefont {Murphy}},\ }\href {\doibase 10.1016/j.sna.2011.05.045} {\bibfield  {journal} {\bibinfo  {journal} {Sensors and Actuators A: Physical}\ }\textbf {\bibinfo {volume} {164}},\ \bibinfo {pages} {104} (\bibinfo {year} {2011})}\BibitemShut {NoStop}%
\bibitem [{\citenamefont {Choi}\ and\ \citenamefont {Kuchenbecker}(2013)}]{choi2013vibrotactile}%
  \BibitemOpen
  \bibfield  {author} {\bibinfo {author} {\bibfnamefont {S.}~\bibnamefont {Choi}}\ and\ \bibinfo {author} {\bibfnamefont {K.~J.}\ \bibnamefont {Kuchenbecker}},\ }\href {\doibase 10.1109/TOH.2012.72} {\bibfield  {journal} {\bibinfo  {journal} {IEEE Transactions on Haptics}\ }\textbf {\bibinfo {volume} {6}},\ \bibinfo {pages} {388} (\bibinfo {year} {2013})}\BibitemShut {NoStop}%
\bibitem [{\citenamefont {Popovic}\ and\ \citenamefont {Keller}(2005)}]{popovic2005nms}%
  \BibitemOpen
  \bibfield  {author} {\bibinfo {author} {\bibfnamefont {D.~B.}\ \bibnamefont {Popovic}}\ and\ \bibinfo {author} {\bibfnamefont {T.}~\bibnamefont {Keller}},\ }\href {\doibase 10.1080/03091900410001714180} {\bibfield  {journal} {\bibinfo  {journal} {Journal of Medical Engineering \& Technology}\ }\textbf {\bibinfo {volume} {29}},\ \bibinfo {pages} {19} (\bibinfo {year} {2005})}\BibitemShut {NoStop}%
\bibitem [{\citenamefont {Stronks}, \citenamefont {Parker},\ and\ \citenamefont {Barnes}(2016)}]{7470536}%
  \BibitemOpen
  \bibfield  {author} {\bibinfo {author} {\bibfnamefont {H.~C.}\ \bibnamefont {Stronks}}, \bibinfo {author} {\bibfnamefont {D.~J.}\ \bibnamefont {Parker}}, \ and\ \bibinfo {author} {\bibfnamefont {N.}~\bibnamefont {Barnes}},\ }\href {\doibase 10.1109/TOH.2016.2569484} {\bibfield  {journal} {\bibinfo  {journal} {IEEE Transactions on Haptics}\ }\textbf {\bibinfo {volume} {9}},\ \bibinfo {pages} {446} (\bibinfo {year} {2016})}\BibitemShut {NoStop}%
\bibitem [{\citenamefont {Barathi~Kanna}\ \emph {et~al.}(2021)\citenamefont {Barathi~Kanna}, \citenamefont {Ganesh~Kumar}, \citenamefont {Niranjan}, \citenamefont {Prashanth}, \citenamefont {Rolant~Gini},\ and\ \citenamefont {Harikumar}}]{9442056}%
  \BibitemOpen
  \bibfield  {author} {\bibinfo {author} {\bibfnamefont {S.}~\bibnamefont {Barathi~Kanna}}, \bibinfo {author} {\bibfnamefont {T.~R.}\ \bibnamefont {Ganesh~Kumar}}, \bibinfo {author} {\bibfnamefont {C.}~\bibnamefont {Niranjan}}, \bibinfo {author} {\bibfnamefont {S.}~\bibnamefont {Prashanth}}, \bibinfo {author} {\bibfnamefont {J.}~\bibnamefont {Rolant~Gini}}, \ and\ \bibinfo {author} {\bibfnamefont {M.}~\bibnamefont {Harikumar}},\ }in\ \href {\doibase 10.1109/ICACCS51430.2021.9442056} {\emph {\bibinfo {booktitle} {2021 7th International Conference on Advanced Computing and Communication Systems (ICACCS)}}},\ Vol.~\bibinfo {volume} {1}\ (\bibinfo {year} {2021})\ pp.\ \bibinfo {pages} {466--471}\BibitemShut {NoStop}%
\bibitem [{\citenamefont {Ryu}\ \emph {et~al.}(2007)\citenamefont {Ryu}, \citenamefont {Jung}, \citenamefont {Kim},\ and\ \citenamefont {Choi}}]{4415100}%
  \BibitemOpen
  \bibfield  {author} {\bibinfo {author} {\bibfnamefont {J.}~\bibnamefont {Ryu}}, \bibinfo {author} {\bibfnamefont {J.}~\bibnamefont {Jung}}, \bibinfo {author} {\bibfnamefont {S.}~\bibnamefont {Kim}}, \ and\ \bibinfo {author} {\bibfnamefont {S.}~\bibnamefont {Choi}},\ }in\ \href {\doibase 10.1109/ROMAN.2007.4415100} {\emph {\bibinfo {booktitle} {RO-MAN 2007 - The 16th IEEE International Symposium on Robot and Human Interactive Communication}}}\ (\bibinfo {year} {2007})\ pp.\ \bibinfo {pages} {310--315}\BibitemShut {NoStop}%
\bibitem [{\citenamefont {Muthuselvi}\ \emph {et~al.}(2022)\citenamefont {Muthuselvi}, \citenamefont {Bharathi}, \citenamefont {Gladina~Sherby}, \citenamefont {Jeya~Monica},\ and\ \citenamefont {Reshma}}]{10010851}%
  \BibitemOpen
  \bibfield  {author} {\bibinfo {author} {\bibfnamefont {M.}~\bibnamefont {Muthuselvi}}, \bibinfo {author} {\bibfnamefont {R.}~\bibnamefont {Bharathi}}, \bibinfo {author} {\bibfnamefont {M.}~\bibnamefont {Gladina~Sherby}}, \bibinfo {author} {\bibfnamefont {J.}~\bibnamefont {Jeya~Monica}}, \ and\ \bibinfo {author} {\bibfnamefont {S.}~\bibnamefont {Reshma}},\ }in\ \href {\doibase 10.1109/ICAISS55157.2022.10010851} {\emph {\bibinfo {booktitle} {2022 International Conference on Augmented Intelligence and Sustainable Systems (ICAISS)}}}\ (\bibinfo {year} {2022})\ pp.\ \bibinfo {pages} {689--694}\BibitemShut {NoStop}%
\bibitem [{\citenamefont {Lien}, \citenamefont {Chang},\ and\ \citenamefont {Chen}(2020)}]{lien2020foam}%
  \BibitemOpen
  \bibfield  {author} {\bibinfo {author} {\bibfnamefont {Y.-H.}\ \bibnamefont {Lien}}, \bibinfo {author} {\bibfnamefont {S.-C.}\ \bibnamefont {Chang}}, \ and\ \bibinfo {author} {\bibfnamefont {Y.-P.}\ \bibnamefont {Chen}},\ }\href {\doibase 10.1016/j.jbmt.2020.07.015} {\bibfield  {journal} {\bibinfo  {journal} {Journal of Bodywork and Movement Therapies}\ }\textbf {\bibinfo {volume} {24}},\ \bibinfo {pages} {202} (\bibinfo {year} {2020})}\BibitemShut {NoStop}%
\bibitem [{\citenamefont {Gursky}\ \emph {et~al.}(2022)\citenamefont {Gursky}, \citenamefont {Krot}, \citenamefont {Korendiy},\ and\ \citenamefont {Zimroz}}]{Gursky2022}%
  \BibitemOpen
  \bibfield  {author} {\bibinfo {author} {\bibfnamefont {V.}~\bibnamefont {Gursky}}, \bibinfo {author} {\bibfnamefont {P.}~\bibnamefont {Krot}}, \bibinfo {author} {\bibfnamefont {V.}~\bibnamefont {Korendiy}}, \ and\ \bibinfo {author} {\bibfnamefont {R.}~\bibnamefont {Zimroz}},\ }\href {\doibase 10.3390/machines10020130} {\bibfield  {journal} {\bibinfo  {journal} {Machines}\ }\textbf {\bibinfo {volume} {10}},\ \bibinfo {pages} {130} (\bibinfo {year} {2022})}\BibitemShut {NoStop}%
\bibitem [{\citenamefont {Wang}\ \emph {et~al.}(2020)\citenamefont {Wang}, \citenamefont {Feng}, \citenamefont {Huang}, \citenamefont {Fang},\ and\ \citenamefont {Wang}}]{mi11080753}%
  \BibitemOpen
  \bibfield  {author} {\bibinfo {author} {\bibfnamefont {R.}~\bibnamefont {Wang}}, \bibinfo {author} {\bibfnamefont {Z.}~\bibnamefont {Feng}}, \bibinfo {author} {\bibfnamefont {S.}~\bibnamefont {Huang}}, \bibinfo {author} {\bibfnamefont {X.}~\bibnamefont {Fang}}, \ and\ \bibinfo {author} {\bibfnamefont {J.}~\bibnamefont {Wang}},\ }\href {\doibase 10.3390/mi11080753} {\bibfield  {journal} {\bibinfo  {journal} {Micromachines}\ }\textbf {\bibinfo {volume} {11}} (\bibinfo {year} {2020}),\ 10.3390/mi11080753}\BibitemShut {NoStop}%
\bibitem [{\citenamefont {Won}\ and\ \citenamefont {Lee}(2005)}]{1519521}%
  \BibitemOpen
  \bibfield  {author} {\bibinfo {author} {\bibfnamefont {S.~H.}\ \bibnamefont {Won}}\ and\ \bibinfo {author} {\bibfnamefont {J.}~\bibnamefont {Lee}},\ }\href {\doibase 10.1109/TMAG.2005.855152} {\bibfield  {journal} {\bibinfo  {journal} {IEEE Transactions on Magnetics}\ }\textbf {\bibinfo {volume} {41}},\ \bibinfo {pages} {4018} (\bibinfo {year} {2005})}\BibitemShut {NoStop}%
\bibitem [{\citenamefont {Chen}(2013)}]{Chen2013}%
  \BibitemOpen
  \bibfield  {author} {\bibinfo {author} {\bibfnamefont {Y.}~\bibnamefont {Chen}},\ }\href@noop {} {\bibfield  {journal} {\bibinfo  {journal} {Michigan State University}\ ,\ \bibinfo {pages} {1}} (\bibinfo {year} {2013})}\BibitemShut {NoStop}%
\bibitem [{\citenamefont {Ganesan}, \citenamefont {Do},\ and\ \citenamefont {Seshia}(2016)}]{ganesan2016observation}%
  \BibitemOpen
  \bibfield  {author} {\bibinfo {author} {\bibfnamefont {A.}~\bibnamefont {Ganesan}}, \bibinfo {author} {\bibfnamefont {C.}~\bibnamefont {Do}}, \ and\ \bibinfo {author} {\bibfnamefont {A.}~\bibnamefont {Seshia}},\ }\href@noop {} {\bibfield  {journal} {\bibinfo  {journal} {Applied Physics Letters}\ }\textbf {\bibinfo {volume} {109}} (\bibinfo {year} {2016})}\BibitemShut {NoStop}%
\bibitem [{\citenamefont {Ganesan}, \citenamefont {Do},\ and\ \citenamefont {Seshia}(2017{\natexlab{a}})}]{ganesan2017excitation}%
  \BibitemOpen
  \bibfield  {author} {\bibinfo {author} {\bibfnamefont {A.}~\bibnamefont {Ganesan}}, \bibinfo {author} {\bibfnamefont {C.}~\bibnamefont {Do}}, \ and\ \bibinfo {author} {\bibfnamefont {A.}~\bibnamefont {Seshia}},\ }\href@noop {} {\bibfield  {journal} {\bibinfo  {journal} {Europhysics Letters}\ }\textbf {\bibinfo {volume} {119}},\ \bibinfo {pages} {10002} (\bibinfo {year} {2017}{\natexlab{a}})}\BibitemShut {NoStop}%
\bibitem [{\citenamefont {Ganesan}, \citenamefont {Do},\ and\ \citenamefont {Seshia}(2017{\natexlab{b}})}]{ganesan2017phononic}%
  \BibitemOpen
  \bibfield  {author} {\bibinfo {author} {\bibfnamefont {A.}~\bibnamefont {Ganesan}}, \bibinfo {author} {\bibfnamefont {C.}~\bibnamefont {Do}}, \ and\ \bibinfo {author} {\bibfnamefont {A.}~\bibnamefont {Seshia}},\ }\href@noop {} {\bibfield  {journal} {\bibinfo  {journal} {Physical review letters}\ }\textbf {\bibinfo {volume} {118}},\ \bibinfo {pages} {033903} (\bibinfo {year} {2017}{\natexlab{b}})}\BibitemShut {NoStop}%
\end{thebibliography}%

\end{document}